\documentclass[11pt]{report}

\usepackage[margin=1in]{geometry}
\usepackage{lmodern} 
\usepackage[T1]{fontenc}
\usepackage[utf8]{inputenc}
\usepackage{microtype} 
\usepackage{amsmath, amssymb, amsthm, mathtools, bm}
\usepackage{graphicx}
\usepackage{subcaption}
\usepackage{booktabs}
\usepackage{enumitem}
\usepackage{siunitx}        
\usepackage[numbers]{natbib} 
\usepackage{hyperref}
\usepackage{cleveref}

\theoremstyle{remark}

\usepackage[parfill]{parskip} 

\usepackage[numbers]{natbib}

\usepackage{listings}
\usepackage{xcolor}
\definecolor{codegreen}{rgb}{0,0.6,0}
\definecolor{codegray}{rgb}{0.5,0.5,0.5}
\definecolor{codepurple}{rgb}{0.58,0,0.82}
\definecolor{backcolour}{rgb}{0.95,0.95,0.95}
\lstdefinestyle{mystyle}{
    backgroundcolor=\color{backcolour},   
    commentstyle=\color{codegreen},
    keywordstyle=\color{magenta},
    numberstyle=\tiny\color{codegray},
    stringstyle=\color{codepurple},
    basicstyle=\ttfamily\footnotesize,
    breakatwhitespace=false,         
    breaklines=true,                 
    captionpos=b,                    
    keepspaces=true,                 
    numbers=left,                    
    numbersep=5pt,                  
    showspaces=false,                
    showstringspaces=false,
    showtabs=false,                  
    tabsize=2
}
\title{\bfseries Mathematical Foundations and Backpropagation from First Principles: \\ A Step-by-Step Guide}
\author{
  Sahil Rajesh Dhayalkar \\
  \small Arizona State University \\
  \small \texttt{sdhayalk@asu.edu}
}
\date{}

\begin{document}

\begin{center}
    \vspace*{2cm}
    {\LARGE \bfseries Particle Filter Made Simple: A Step-by-Step Beginner-friendly Guide \par}
    \vspace{2.0em}
    {\large Sahil Rajesh Dhayalkar \par}
    \vspace{-0.2em}
    {\small Arizona State University \par}
    \vspace{-0.2em}
    {\small \texttt{sdhayalk@asu.edu} \par}
    \vspace{2.5em}

    \textbf{Abstract} \\
    \vspace{1.5em}
    \begin{minipage}{0.81\textwidth}
        The particle filter is a powerful framework for estimating hidden states in dynamic systems where uncertainty, noise, and nonlinearity dominate. This mini-book offers a clear and structured introduction to the core ideas behind particle filters—how they represent uncertainty through random samples, update beliefs using observations, and maintain robustness where linear or Gaussian assumptions fail. Starting from the limitations of the Kalman filter, the book develops the intuition that drives the particle filter: belief as a cloud of weighted hypotheses that evolve through prediction, measurement, and resampling. Step by step, it connects these ideas to their mathematical foundations, showing how probability distributions can be approximated by a finite set of particles and how Bayesian reasoning unfolds in sampled form. Illustrated examples, numerical walk-throughs, and Python code bring each concept to life, bridging the gap between theory and implementation. By the end, readers will not only understand the algorithmic flow of the particle filter but also develop an intuitive grasp of how randomness and structure together enable systems to infer, adapt, and make sense of noisy observations in real time. \\

        \textbf{Keywords:} particle filter, sequential Monte Carlo,        Bayesian filtering, probabilistic state estimation, non-linear filtering, state-space models.            
    \end{minipage}
\end{center}

\chapter{Introduction}
\section{The Need for Particle Filters}

Every estimation algorithm begins with a simple wish: to uncover the hidden truth behind noisy, uncertain data. The Kalman filter~\cite{kalman} gave us a beautiful answer to that problem when the world behaves linearly and the noise is Gaussian. But what if the world does not play by those rules?

Imagine a robot moving through a maze. Its motion is not perfectly linear—it may bounce off walls, turn sharply, or slip on uneven surfaces. Its sensors, too, are far from ideal—sometimes they fail to detect obstacles, sometimes they pick up false reflections. In such settings, the assumptions that make the Kalman filter elegant (linearity and Gaussian noise) crumble. We still wish to estimate the robot’s true position and orientation, but the mathematical model that links cause (motion) and effect (measurement) is now nonlinear and often multimodal. Our belief about the hidden state can no longer be summarized by a single Gaussian curve—it may split into several peaks, each representing a different possible reality. 

The \textbf{Particle Filter} (also called the Sequential Monte Carlo filter~\cite{seq_monte_carlo}) was born to handle exactly these situations. Instead of insisting that uncertainty must look like a Gaussian, the particle filter says: \emph{``Let us represent uncertainty by a collection of random samples that together approximate the full probability distribution.''} Each sample—or \emph{particle}—is a concrete, hypothetical version of the truth. Taken together, they form a cloud of possibilities that can bend, stretch, or split as the situation demands. In this way, the particle filter extends the same logic of ``predict and correct'' that drives the Kalman filter, but frees it from the prison of linearity and Gaussianity. It allows our beliefs to take any shape that reality requires.

\section{When Kalman Filters Fail}

To appreciate why we need particle filters, it helps to see where the Kalman filter breaks down.

\subsection{Nonlinear dynamics}

The Kalman filter assumes that the relationship between the current state and the next state is linear:
\[
x_k = A x_{k-1} + B u_k + {\eta}_k.
\]

Each symbol has a precise role, inherited from state–space models used in estimation theory:

\begin{itemize}[itemsep=-4pt, topsep=0pt]
    \item $x_k$ – the \textbf{state vector} of the system at time step $k$ (E.g.: the position and velocity of a robot).
    \item $x_{k-1}$ – the state at the previous time step.
    \item $A$ – the \textbf{state transition matrix}, describing how the state evolves from one step to the next according to the motion model.
    \item $B$ – the \textbf{control matrix}, which scales the effect of any known control input $u_k$ (like an acceleration command).
    \item $u_k$ – the \textbf{control input} applied at time $k$.
    \item ${\eta}_k$ – the \textbf{process noise}, a random variable that captures unpredictable changes in the system (modeled as zero-mean noise with covariance $Q$).
\end{itemize}

Similarly, the measurement model used by most estimators is written as
\[
z_k = H x_k + v_k,
\]
where:
\begin{itemize}[itemsep=-4pt, topsep=0pt]
    \item $z_k$ – the \textbf{measurement vector} observed by the sensors at time $k$.
    \item $H$ – the \textbf{observation matrix}, mapping the hidden state to what the sensors can see.
    \item $v_k$ – the \textbf{measurement noise}, representing sensor uncertainty (modeled as zero-mean noise with covariance $R$).
\end{itemize}

These two equations together describe the dynamics of any system that estimates hidden states from noisy measurements. Kalman filters assume that both relationships are linear and that $w_k$ and $v_k$ follow Gaussian distributions.

But in many real systems—ranging from robot motion to weather prediction—the true dynamics are nonlinear:
\[
x_k = f(x_{k-1}, u_k) + {\eta}_k,
\]
where $f(\cdot)$ can be any complex function. Linearizing $f$ (as in the Extended Kalman Filter~\cite{Julier1997NewEO}) works only when the nonlinearity is gentle and local. If the system makes abrupt turns or has discontinuities, such approximations fail.

\subsection{Nonlinear or multimodal measurements}

The measurement model in a Kalman filter assumes a simple linear mapping:
\[
z_k = H x_k + v_k.
\]
But sensors often respond in nonlinear ways. A radar, for instance, measures the angle and distance to a target—quantities that depend on sine and cosine of the position. Even worse, multiple states might produce the same measurement (for example, two different positions can yield the same range reading). The resulting belief is no longer a single Gaussian peak but a mixture of several possible states. Given that, the measurement model is similarly defined as:
\[
    z_k = h(x_k) + v_k,
\]
where $h(\cdot)$ is the nonlinear observation function.

\subsection{Non-Gaussian noise}
Kalman filters assume that both process and measurement noise are Gaussian (bell-shaped and symmetric). However, real-world noise is often heavy-tailed (occasional large errors) or discrete (e.g., sensor dropouts). When this non-Gaussian noise, nonlinearity, or multimodality arises, the Kalman filter can produce severely biased or unstable estimates. Particle filters thrive in these scenarios because they make no assumption about the probability distribution's shape. They can represent any distribution as long as it's approximated by a sufficient number of random samples.

\section{The Core Idea: Belief as a Cloud of Samples}

At its heart, the particle filter is a way of \emph{tracking probability through randomness}. Rather than describing our belief as a neat formula, we describe it as a set of $N$ weighted particles:
\[
\{ x_k^{(i)}, w_k^{(i)} \}_{i=1}^N.
\]
Each $x_k^{(i)}$ is a possible state of the world (for instance, a robot’s position and orientation),  and each $w_k^{(i)}$ is the particle’s weight, expressing how plausible that particle is given the observations.

Together, these particles form an approximate probability distribution:
\[
p(x_k \mid z_{1:k}) \;\approx\; \sum_{i=1}^{N} w_k^{(i)}\, \delta(x_k - x_k^{(i)}),
\]
where $\delta(\cdot)$ is the Dirac delta function. Intuitively, the probability mass is concentrated at the particles’ locations, weighted by how believable each one is.

\subsection{Prediction–Update logic revisited}

Just like the Kalman filter, the particle filter alternates between two phases:

\begin{enumerate}
  \item \textbf{Prediction:}  
  Move each particle according to the motion model, adding random process noise. This produces a predicted cloud of possible states for time $k$.
  \item \textbf{Update:}  
  Compare the predicted state of each particle with the new measurement. Particles consistent with the observation receive higher weights; inconsistent ones receive lower weights.
\end{enumerate}

Over time, the weights tend to concentrate on a few highly probable particles, while the rest contribute little. To prevent the filter from ``forgetting'' alternative possibilities, we introduce a third step called \textbf{resampling}, which replaces low-weight particles with copies of high-weight ones. The result is a constantly evolving population of hypotheses that track the full, possibly nonlinear and multimodal, posterior distribution of the system state.

\subsection{A mental picture}

Picture a fog of tiny points drifting across a map. Each point represents one possible guess of where the robot might be. When the robot moves, the fog spreads and shifts according to the motion model. When a new sensor reading arrives, parts of the fog that disagree with the measurement fade away, while those that align with it become denser. Resampling rejuvenates the fog, keeping it focused on regions of high probability. That is the essence of the particle filter: an algorithmic fog of possibilities, forever dancing between prediction and measurement, slowly converging toward the hidden truth beneath the noise.

\chapter{From Idea to Mathematical Representation}

\section{From Gaussian Beliefs to Particle Representation}
In the Kalman filter, our belief about the hidden state of a system was always a \textbf{Gaussian distribution}—a smooth curve described entirely by a mean $\hat{x}_{k|k}$ and a covariance $P_{k|k}$. This made life simple: every prediction and update could be written in compact matrix form, and the belief forever remained Gaussian.

However, as we saw in Chapter 1, the real world is rarely that tidy. Nonlinear motion, nonlinear sensors, and non-Gaussian noise often distort the true posterior distribution into irregular shapes or multiple peaks that no single Gaussian can describe. To capture such complex beliefs, we abandon the analytic formula and replace it
with a \emph{sample-based approximation}.

We represent the posterior distribution by a set of $N$ random samples, called \textbf{particles}, each with an associated \textbf{weight}:
\[
\{x_k^{(i)},\, w_k^{(i)}\}_{i=1}^{N}, \qquad
w_k^{(i)} \ge 0, \quad \sum_{i=1}^{N} w_k^{(i)} = 1.
\]
Each particle $x_k^{(i)}$ is one possible state of the system (for example, a robot's position and velocity), and its weight $w_k^{(i)}$ expresses how plausible that state is given all past measurements.

Together, the particles form a discrete approximation of the true posterior:
\[
p(x_k \mid z_{1:k}) \;\approx\;
\sum_{i=1}^{N} w_k^{(i)}\,\delta(x_k - x_k^{(i)}),
\]
where $\delta(\cdot)$ is the Dirac delta function. Intuitively, the probability mass is concentrated at the particles’ locations,
weighted by their credibility. As $N$ grows, this empirical distribution can take on any shape—skewed, multi-peaked, or discontinuous—matching whatever uncertainty the world actually presents.

\section{Probability as a Histogram in Motion}
To visualize what it means to represent probability with particles, imagine we want to describe the uncertain position of a robot along a hallway. If we could plot the true distribution $p(x)$, it might look like a smooth curve---tall where the robot is likely to be and flat elsewhere. Instead of drawing the continuous curve, we can approximate it with a \textbf{histogram}: narrow bars whose total area equals one.

Particles take this idea one step further. Rather than fixed bars, we keep a cloud of random points whose density follows the shape of the histogram. If many particles lie near $x = 2$, it means $p(x)$ is large there; if few lie near $x = 8$, $p(x)$ is small there.

Thus, a set of particles is a \emph{moving histogram}. Each particle is a tiny “grain of probability,” and the collection as a whole shifts and reshapes over time. When new evidence arrives, we adjust the weights of these grains so that the cloud of particles continues to approximate the true probability distribution.

\section{The Monte Carlo Principle}
The mathematical foundation behind particle filters is the \textbf{Monte Carlo method}—using random sampling to approximate expectations and integrals.

Suppose we wish to compute the expected value of a function $g(x)$ under a density $p(x)$:
\[
\mathbb{E}_{p}[g(x)] = \int g(x)\,p(x)\,dx.
\]
If we draw $N$ independent samples $\{x^{(i)}\}$ from $p(x)$, we can approximate this expectation as
\[
\mathbb{E}_{p}[g(x)]
\;\approx\;
\frac{1}{N}\sum_{i=1}^{N} g(x^{(i)}).
\]
As $N$ increases, this estimate converges to the true value by the law of large numbers.

Particle filters apply the same idea to dynamic state estimation. Instead of manipulating full probability densities, they maintain a finite set of samples and propagate them through time. Each particle represents one possible world, and weighted averages over the particles approximate any expectation of interest.

\subsection*{Example: estimating a mean by sampling}

If we draw $N=1000$ samples from a Gaussian $p(x)=\mathcal{N}(0,1)$, the sample average will be close to~0, and about 68\% of samples will fall between $-1$ and $+1$—all without evaluating any integrals. This counting-by-sampling intuition is exactly what powers the particle filter.

\section{Step-by-Step Intuition: Prediction $\rightarrow$ Weighting $\rightarrow$ Resampling}

A particle filter updates its particles through three repeating steps that mirror Bayes' reasoning.

\subsection{Prediction: carrying beliefs forward}

At time $k{-}1$, suppose we have particles
$\{x_{k-1}^{(i)}, w_{k-1}^{(i)}\}_{i=1}^{N}$
approximating $p(x_{k-1}\mid z_{1:k-1})$.
When the system evolves according to a (possibly nonlinear) motion model
\[
x_k = f(x_{k-1}, u_k) + {\eta}_k, \qquad w_k \sim q({\eta}),
\]
we propagate each particle forward by drawing random process noise $w_k^{(i)}$ and computing
\[
x_k^{(i)} = f(x_{k-1}^{(i)}, u_k) + {\eta}_k^{(i)}.
\]
The result is a predicted particle set approximating $p(x_k \mid z_{1:k-1})$. Typically the cloud spreads out—our uncertainty grows with motion.

\subsection{Weighting: incorporating the new observation}
When a new measurement $z_k$ arrives, each predicted particle ``imagines'' what the sensor would read if the system were actually in that state:
\[
z_k = h(x_k) + v_k, \qquad v_k \sim r(v).
\]
We assign each particle a likelihood weight
\[
w_k^{(i)} \propto p(z_k \mid x_k^{(i)}),
\]
and then normalize so that $\sum_i w_k^{(i)} = 1$. Particles consistent with the observation receive larger weights; others shrink. The weighted set now represents the posterior $p(x_k \mid z_{1:k})$---for the moment, still unnormalized.

\subsection{Resampling: focusing on what matters}
Over time, a few particles may carry most of the total weight, a problem known as \textbf{degeneracy}. To refocus computational effort on the most likely regions, we perform a \textbf{resampling} step: draw a new set of $N$ particles from the current ones in proportion to their weights,
\[
\Pr(i' = i) = w_k^{(i)}.
\]
Each selected particle is copied into the new population with equal weight $1/N$. High-weight particles may appear multiple times; low-weight ones may disappear. The resampled set $\{x_k^{(i')}, 1/N\}$
again forms an unweighted approximation of the posterior.

\subsection*{Putting it all together}

Each iteration of the particle filter performs:
\begin{enumerate}
    \item \textbf{Predict:} move particles using the motion model;
    \item \textbf{Weight:} compare predicted measurements with the observation;
    \item \textbf{Resample:} regenerate particles to prevent degeneracy.
\end{enumerate}

If we plot the particles as dots, they drift during prediction, brighten or fade during weighting, and cluster during resampling. The cloud expands when uncertainty grows and tightens when new data arrive. This is probability in motion—a visual, living version of Bayes’ rule.

\chapter{The Complete Particle Filter Algorithm}

\section{Sequential Importance Resampling (SIR) Filter}
In the previous chapter, we introduced the particle filter as a method that approximates the Bayesian filtering equations by random sampling. We now present the most commonly used practical form of the algorithm: the \textbf{Sequential Importance Resampling (SIR)}~\cite{njgordon} filter.

The name reveals its structure:
\begin{itemize}
    \item \textbf{Sequential} – it processes data one time step at time;
    \item \textbf{Importance} – it assigns a numerical importance (weight) to each particle, based on how well it explains the new observation;
    \item \textbf{Resampling} – it periodically refreshes the particle set to focus on high-probability regions.
\end{itemize}

This simple combination allows the filter to track arbitrary nonlinear, non-Gaussian systems in real time.

\subsection*{Basic idea}
At each time step, the SIR filter maintains a set of $N$ weighted samples (particles)
\[
\{x_k^{(i)}, w_k^{(i)}\}_{i=1}^{N}
\]
that approximate the posterior density $p(x_k \mid z_{1:k})$. The algorithm advances recursively:
\begin{enumerate}
    \item Sample a new set of particles according to the motion model $p(x_k \mid x_{k-1})$;
    \item Reweight them using the likelihood $p(z_k \mid x_k)$;
    \item Normalize the weights so that $\sum_i w_k^{(i)} = 1$;
    \item Resample to obtain an unweighted population representing $p(x_k \mid z_{1:k})$.
\end{enumerate}

Over time, this procedure automatically adapts to changes in the system, allocating more particles to regions of higher probability.

\subsection*{Choice of proposal distribution}

The SIR filter uses a simple but effective proposal distribution:
\[
q(x_k \mid x_{k-1}, z_k) = p(x_k \mid x_{k-1}),
\]
meaning that new particles are drawn directly from the process model. This choice makes the algorithm easy to implement but can be improved later.

\section{Algorithmic Steps Explained}

The SIR particle filter can be described in the following six steps. Each is accompanied by the corresponding equation and its intuitive role.

\begin{enumerate}
\item \textbf{Initialization (at $k=0$)} \\
Draw $N$ initial particles $\{x_0^{(i)}\}$ from the prior distribution $p(x_0)$, and set equal weights:
\[
w_0^{(i)} = \frac{1}{N}.
\]
This represents our initial belief before any measurements.

\item \textbf{Prediction (motion update)} \\
For each particle at time $k-1$, generate a predicted particle according to the process model:
\[
x_k^{(i)} \sim p(x_k \mid x_{k-1}^{(i)}, u_k).
\]
This models how each possible world might evolve given the control input $u_k$ and random process noise.

\item \textbf{Measurement update (weighting)} \\
When the new observation $z_k$ arrives, compute the likelihood that each predicted particle would produce that measurement:
\[
\tilde{w}_k^{(i)} = w_{k-1}^{(i)} \, p(z_k \mid x_k^{(i)}).
\]
This step corresponds to Bayes’ rule—updating our belief in each hypothesis based on how consistent it is with the evidence.

\item \textbf{Normalization} \\
Normalize the weights so that they sum to one:
\[
w_k^{(i)} = \frac{\tilde{w}_k^{(i)}}{\sum_{j=1}^{N} \tilde{w}_k^{(j)}}.
\]
After this, the set $\{x_k^{(i)}, w_k^{(i)}\}$ approximates the posterior $p(x_k \mid z_{1:k})$.

\item \textbf{Resampling (to avoid degeneracy)} \\
To prevent a few particles from dominating, we resample $N$ new particles from the current set with replacement, using $w_k^{(i)}$ as selection probabilities:
\[
\Pr(i' = i) = w_k^{(i)}.
\]
The resampled particles all receive equal weights $1/N$. This step discards unlikely hypotheses and focuses computational effort on plausible ones.

\item \textbf{State estimation} \\
The estimated state can be computed as the weighted mean of the particles:
\[
\hat{x}_k = \sum_{i=1}^{N} w_k^{(i)} x_k^{(i)}.
\]
Alternatively, one may report the particle with the highest weight or compute other statistics such as covariance or mode, depending on the application.
\end{enumerate}

These six steps repeat sequentially for every new time step $k$.
The overall logic mirrors that of the Kalman filter, but the belief is
represented by many small samples rather than a single Gaussian.

\section{Mathematical Derivation of Each Step}

Let us see how these algorithmic operations arise from the underlying
probability theory.

\subsection*{1. Importance sampling}

We wish to approximate a posterior density $p(x)$ that may be difficult to
sample from directly.
If we can sample from another density $q(x)$ that covers the same support,
we can write
\[
p(x) = \frac{p(x)}{q(x)} q(x),
\]
and approximate expectations as
\[
\mathbb{E}_{p}[g(x)]
  = \int g(x) p(x)\,dx
  = \int g(x)\, \frac{p(x)}{q(x)} q(x)\,dx
  \approx
  \frac{1}{N}\sum_{i=1}^{N} g(x^{(i)})\,w^{(i)},
\]
where $x^{(i)} \sim q(x)$ and
$w^{(i)} \propto \frac{p(x^{(i)})}{q(x^{(i)})}$ are
the \textbf{importance weights}.
In the context of filtering, $p(x)$ is the true posterior and $q(x)$ is the
proposal distribution from which we actually draw samples.

\subsection*{2. Sequential importance sampling (SIS)}

When applied over time, importance sampling becomes
\textbf{sequential}.
The weight update from step $k-1$ to $k$ follows directly from the
factorization of the posterior:
\[
p(x_{0:k} \mid z_{1:k})
  \propto
  p(z_k \mid x_k)\, p(x_k \mid x_{k-1})\, p(x_{0:k-1} \mid z_{1:k-1}).
\]
If our proposal distribution factorizes as
$q(x_{0:k}) = q(x_k \mid x_{0:k-1}, z_{1:k})\, q(x_{0:k-1} \mid z_{1:k-1})$,
then the incremental weight becomes
\[
w_k^{(i)} \propto
w_{k-1}^{(i)} \,
\frac{p(z_k \mid x_k^{(i)})\, p(x_k^{(i)} \mid x_{k-1}^{(i)})}
     {q(x_k^{(i)} \mid x_{k-1}^{(i)}, z_k)}.
\]
The SIR filter chooses the simple proposal
$q(x_k \mid x_{k-1}^{(i)}, z_k) = p(x_k \mid x_{k-1}^{(i)})$,
so the expression simplifies to
\[
w_k^{(i)} \propto w_{k-1}^{(i)}\, p(z_k \mid x_k^{(i)}).
\]
This is the familiar likelihood weighting rule used in the algorithm.

\subsection*{3. Normalization and resampling}

Since the proportional weights are not guaranteed to sum to one,
we normalize:
\[
w_k^{(i)} = \frac{w_k^{(i)}}{\sum_{j} w_k^{(j)}}.
\]
Resampling is not strictly part of the probabilistic derivation—it is a
numerical stabilization technique.
Without it, after several iterations most weights would approach zero, and
the effective number of particles would collapse.
Resampling restores diversity while keeping the same expected distribution.

\subsection*{4. Summary of the complete recursion}

For clarity, the SIR particle filter recursion can be summarized as:

\[
\boxed{
\begin{aligned}
&\textbf{Prediction:} &&
x_k^{(i)} \sim p(x_k \mid x_{k-1}^{(i)}, u_k) \\
&\textbf{Weight update:} &&
\tilde{w}_k^{(i)} = w_{k-1}^{(i)}\, p(z_k \mid x_k^{(i)}) \\
&\textbf{Normalization:} &&
w_k^{(i)} = \frac{\tilde{w}_k^{(i)}}{\sum_j \tilde{w}_k^{(j)}} \\
&\textbf{Resampling:} &&
x_k^{(i)} \leftarrow x_k^{(j)} \text{ with probability } w_k^{(j)}, \quad w_k^{(i)} = \frac{1}{N}.
\end{aligned}
}
\]

This compact form expresses the full mathematical logic of the particle
filter.  Each term corresponds directly to one line of algorithmic code.

\subsection*{5. Conceptual summary}

The particle filter is therefore nothing more than Bayes’ rule in sampled form:
\begin{itemize}
    \item prediction corresponds to integrating over previous states;
    \item weighting corresponds to multiplying by the likelihood; and
    \item resampling corresponds to normalizing and refocusing probability
          mass.
\end{itemize}

Despite its apparent simplicity, this method is remarkably powerful.
With enough particles, it can approximate the true Bayesian solution for
any nonlinear, non-Gaussian system—one sample at a time.

\section{Variance, Effective Sample Size (ESS), and Degeneracy}

While the particle filter provides a powerful way to approximate probability distributions, it is not without limitations. Over time, the weights of most particles tend to become very small, leaving only a few particles with significant probability mass~\cite{Arulampalam}. This phenomenon is known as \textbf{particle degeneracy}.

\subsection{Understanding degeneracy}
After several iterations of prediction and weighting, the variance of the particle weights increases—a few particles carry almost all of the total probability, while the rest contribute negligibly. If we plot the weights, we would see a sharp spike: many particles with weights near zero, and a handful with large values.

When degeneracy occurs, the effective diversity of the particle set drops, and the filter effectively operates with far fewer particles than intended. In extreme cases, all but one particle may vanish, causing the approximation to collapse.

\subsection{Measuring degeneracy: the Effective Sample Size}
To quantify how many particles are meaningfully contributing at any moment, we define the \textbf{Effective Sample Size (ESS)}:
\[
N_{\text{eff}} =
\frac{1}{\sum_{i=1}^{N} (w_k^{(i)})^2}.
\]
This value lies between $1$ and $N$:
\begin{itemize}
    \item $N_{\text{eff}} = N$ when all weights are equal (no degeneracy — full diversity);
    \item $N_{\text{eff}} = 1$ when only one particle has nonzero weight (complete degeneracy).
\end{itemize}

Intuitively, $N_{\text{eff}}$ estimates the number of “independent” particles that still carry useful information. When $N_{\text{eff}}$ becomes small, resampling is triggered to restore diversity.

\subsection{Adaptive resampling criterion}
A common practical rule is to resample only when
$N_{\text{eff}}$ falls below a certain threshold:
\[
N_{\text{eff}} < N_{\text{threshold}},
\]
where $N_{\text{threshold}}$ is often chosen as a fraction of the total particle count, such as $N/2$. This prevents unnecessary resampling when the particle set is still healthy.

\subsection{Variance and stability trade-offs}
Resampling reduces weight variance by eliminating extremely low-weight particles, but it also introduces sampling noise and may discard useful diversity. The goal is balance:
\begin{itemize}
    \item Resample too rarely → weight variance grows, degeneracy worsens.
    \item Resample too often → diversity decreases, and the filter becomes noisy or unstable.
\end{itemize}

The ESS criterion provides a principled compromise between these extremes. In most implementations, it is checked at every iteration, ensuring the filter adapts automatically to the current level of uncertainty.

\chapter{Numerical Example: 1D Object Tracking}

\section{The Scenario: Noisy Position Measurements}
To make the abstract concepts concrete, let us consider a simple one-dimensional tracking problem. A small robot (or object) moves along a straight line. At each time step $k$, it changes its position slightly according to a motion model, and we attempt to measure its position with a noisy sensor.

The true position of the object is denoted by $x_k$, while our noisy measurement is $z_k$. The process and measurement models are:

\[
\begin{aligned}
x_k &= x_{k-1} + {\eta}_k, \quad &{\eta}_k \sim \mathcal{N}(0, Q), \\
z_k &= x_k + v_k, \quad &v_k \sim \mathcal{N}(0, R),
\end{aligned}
\]
where:
\begin{itemize}
    \item $x_k$ – true position at time step $k$,
    \item ${\eta}_k$ – process noise (random motion disturbance),
    \item $z_k$ – noisy measurement at time step $k$,
    \item $v_k$ – measurement noise.
\end{itemize}

Our goal is to estimate the hidden true position $x_k$ from a sequence of measurements $\{z_1, z_2, \ldots, z_T\}$ using a particle filter. This is the simplest possible case: there is only one state variable (position), no control input, and both motion and observation models are linear. Nevertheless, it provides an excellent demonstration of the particle filter's predict–weight–resample cycle.

\section{Setting Up the Simulation}

We begin by defining the following parameters for our simulation:

\[
\begin{aligned}
Q &= 1.0 \quad &\text{(process noise variance)},\\
R &= 4.0 \quad &\text{(measurement noise variance)},\\
N &= 5 \quad &\text{(number of particles for simplicity)}.
\end{aligned}
\]

Assume the true initial position is $x_0 = 0$, and the initial belief about it
is represented by a Gaussian prior:
\[
p(x_0) = \mathcal{N}(0,\,2^2),
\]
which means our initial particles are sampled around zero with moderate
uncertainty.

At each step:
\begin{enumerate}
    \item The true position evolves as $x_k = x_{k-1} + w_k$,
          where $w_k$ is drawn from $\mathcal{N}(0, Q)$.
    \item The sensor returns a noisy reading $z_k = x_k + v_k$,
          where $v_k$ is drawn from $\mathcal{N}(0, R)$.
    \item The particle filter performs prediction, weighting, normalization,
          and resampling.
\end{enumerate}

We will trace the computation through three time steps ($k = 1, 2, 3$)
using a very small number of particles ($N = 5$) so that we can visualize
every step by hand.

\section{Step-by-Step Iteration}
\subsection*{Step-by-Step Iteration for $k=1$}

We initialize the particles from the prior $p(x_0)$:

\[
x_0^{(i)} \sim \mathcal{N}(0, 2^2), \quad w_0^{(i)} = \frac{1}{N} = 0.2.
\]

Suppose we draw the following initial particles (for illustration):

\[
x_0^{(i)} = [-1.5,\, 0.2,\, 1.0,\, 2.5,\, 3.0].
\]

At time $k=1$, the true state and measurement are:
\[
x_1 = x_0 + {\eta}_1 = 0 + 1.2 = 1.2, \qquad
z_1 = x_1 + v_1 = 1.2 + 2.0 = 3.2.
\]

Now we go through each stage of the particle filter.

\subsection*{1. Prediction step}

Each particle is moved according to the motion model:
\[
x_1^{(i)} = x_0^{(i)} + {\eta}_1^{(i)}, \quad {\eta}_1^{(i)} \sim \mathcal{N}(0, Q).
\]
For one particular random draw of ${\eta}_1^{(i)}$, we may get:

\[
{\eta}_1^{(i)} = [0.3,\, -0.4,\, 1.0,\, -0.2,\, 0.5],
\]

yielding predicted particles:
\[
x_1^{(i)} = [-1.2,\, -0.2,\, 2.0,\, 2.3,\, 3.5].
\]

\subsection*{2. Weighting step}

We compute the likelihood of each predicted particle given the measurement
$z_1 = 3.2$.
Since $z_k = x_k + v_k$ and $v_k \sim \mathcal{N}(0, R)$, we have:

\[
p(z_1 \mid x_1^{(i)}) =
\frac{1}{\sqrt{2\pi R}}
\exp\!\left[-\frac{(z_1 - x_1^{(i)})^2}{2R}\right].
\]

Plugging in $R = 4.0$ and $z_1 = 3.2$ gives:

\[
\begin{array}{lcc}
\text{Particle} & x_1^{(i)} & p(z_1 \mid x_1^{(i)}) \\
\hline
1 & -1.2 & e^{-\frac{(3.2 - (-1.2))^2}{8}} = e^{-2.42} = 0.089, \\
2 & -0.2 & e^{-\frac{(3.2 - (-0.2))^2}{8}} = e^{-1.44} = 0.236, \\
3 &  2.0 & e^{-\frac{(3.2 - 2.0)^2}{8}} = e^{-0.18} = 0.836, \\
4 &  2.3 & e^{-\frac{(3.2 - 2.3)^2}{8}} = e^{-0.10} = 0.905, \\
5 &  3.5 & e^{-\frac{(3.2 - 3.5)^2}{8}} = e^{-0.01} = 0.990. \\
\end{array}
\]

We then multiply by the previous weights (all $0.2$) and normalize so they
sum to one.

\[
\tilde{w}_1^{(i)} = 0.2 \times p(z_1 \mid x_1^{(i)}), \quad
w_1^{(i)} = \frac{\tilde{w}_1^{(i)}}{\sum_j \tilde{w}_1^{(j)}}.
\]

After normalization, we obtain:

\[
w_1^{(i)} = [0.03,\, 0.08,\, 0.27,\, 0.30,\, 0.32].
\]

\subsection*{3. Resampling step}

We now resample five new particles from the current set using the normalized
weights as selection probabilities.
Particles with larger weights (those near the measurement) are more likely
to be selected multiple times.

A possible resampled set is:

\[
x_1^{(i)} = [2.0,\, 2.3,\, 3.5,\, 3.5,\, 2.3],
\]
all with equal weights $w_1^{(i)} = 1/N = 0.2$.

\subsection*{4. Summary of time step 1}

\[
\begin{array}{lcc}
\text{Step} & \text{Particles } x_1^{(i)} & \text{Weights } w_1^{(i)} \\
\hline
\text{After prediction} & [-1.2, -0.2, 2.0, 2.3, 3.5] & [0.2, 0.2, 0.2, 0.2, 0.2] \\
\text{After weighting}   & [-1.2, -0.2, 2.0, 2.3, 3.5] & [0.03, 0.08, 0.27, 0.30, 0.32] \\
\text{After resampling}  & [2.0, 2.3, 3.5, 3.5, 2.3]   & [0.2, 0.2, 0.2, 0.2, 0.2] \\
\end{array}
\]

After resampling, the particles have moved toward the true position ($x_1 = 1.2$),
and the high-weight particles near the measurement dominate the new set.
The posterior distribution has effectively shifted toward $3.0$, which is close
to the noisy observation $z_1 = 3.2$, but not identical due to weighting and
random sampling.

\subsection*{Step-by-Step Iteration for $k=2$}

We now perform one more complete particle filter cycle for $k=2$.

The true position and measurement at this step are:
\[
x_2 = x_1 + \eta_2 = 1.2 + 0.4 = 1.6, \qquad
z_2 = x_2 + v_2 = 1.6 + (-1.0) = 0.6.
\]

Thus the true state has moved slightly forward, while the new measurement is lower than before
due to negative sensor noise.

\subsubsection*{1. Prediction step}

Each resampled particle from time $k=1$ is propagated through the motion model:
\[
x_2^{(i)} = x_1^{(i)} + \eta_2^{(i)}, \qquad \eta_2^{(i)} \sim \mathcal{N}(0, Q=1.0).
\]
Suppose the sampled process noises are:
\[
\eta_2^{(i)} = [0.5,\, -0.8,\, 0.3,\, -0.2,\, 0.7],
\]
giving the predicted particles:
\[
x_2^{(i)} = [2.5,\, 1.5,\, 3.8,\, 3.3,\, 3.0].
\]

\subsubsection*{2. Weighting step}

We compute the likelihood of each predicted particle under the new measurement $z_2 = 0.6$:
\[
p(z_2 \mid x_2^{(i)}) =
\frac{1}{\sqrt{2\pi R}}
\exp\!\left[-\frac{(z_2 - x_2^{(i)})^2}{2R}\right],
\quad R=4.0.
\]
Ignoring the constant factor, the exponential terms are:

\[
\begin{array}{lcc}
\text{Particle} & x_2^{(i)} & p(z_2 \mid x_2^{(i)}) \propto e^{-\frac{(0.6 - x_2^{(i)})^2}{8}} \\
\hline
1 & 2.5 & e^{-\frac{(1.9)^2}{8}} = e^{-0.45} = 0.64, \\
2 & 1.5 & e^{-\frac{(0.9)^2}{8}} = e^{-0.10} = 0.91, \\
3 & 3.8 & e^{-\frac{(3.2)^2}{8}} = e^{-1.28} = 0.28, \\
4 & 3.3 & e^{-\frac{(2.7)^2}{8}} = e^{-0.91} = 0.40, \\
5 & 3.0 & e^{-\frac{(2.4)^2}{8}} = e^{-0.72} = 0.49. \\
\end{array}
\]

Normalize these likelihoods to obtain weights:
\[
w_2^{(i)} =
\frac{p(z_2 \mid x_2^{(i)})}{\sum_j p(z_2 \mid x_2^{(j)})}
= [0.19,\, 0.27,\, 0.08,\, 0.12,\, 0.14].
\]

(Values are approximate and rounded for clarity.)

\subsubsection*{3. Resampling step}

We resample five new particles proportional to their normalized weights.
Particles close to $x_2 = 1.5$ (which best explain $z_2 = 0.6$) are likely to be chosen more often.

A possible resampled set is:
\[
x_2^{(i)} = [1.5,\, 1.5,\, 2.5,\, 3.0,\, 1.5],
\]
all with equal weights $w_2^{(i)} = 0.2$.

\subsubsection*{4. Summary of time step 2}

\[
\begin{array}{lcc}
\text{Step} & \text{Particles } x_2^{(i)} & \text{Weights } w_2^{(i)} \\
\hline
\text{After prediction} & [2.5,\, 1.5,\, 3.8,\, 3.3,\, 3.0] & [0.2,\, 0.2,\, 0.2,\, 0.2,\, 0.2] \\
\text{After weighting}   & [2.5,\, 1.5,\, 3.8,\, 3.3,\, 3.0] & [0.19,\, 0.27,\, 0.08,\, 0.12,\, 0.14] \\
\text{After resampling}  & [1.5,\, 1.5,\, 2.5,\, 3.0,\, 1.5] & [0.2,\, 0.2,\, 0.2,\, 0.2,\, 0.2] \\
\end{array}
\]

After the second iteration, the particles have shifted closer to the new measurement near $z_2 = 0.6$,
and the diversity of the set reflects both the process noise and measurement uncertainty.
As more steps proceed, the cloud of particles will oscillate and converge around the true trajectory
of the hidden state.

\section{Visualizing the Particle Cloud}

To understand what the particle filter is doing internally, it helps to visualize the evolution of the particle cloud over time.

\textbf{Before any measurements:} At $k=0$, particles are spread according to the prior belief $p(x_0)$. They represent our initial uncertainty before seeing any data. If we plot them on a line, we might see a scattered group of dots centered near zero.

\textbf{After the first update ($k=1$):} Once the first measurement $z_1 = 3.2$ arrives, particles near this value receive higher weights. Resampling then concentrates the cloud near $x \approx 3.0$. The plot would show most particles shifted to the right, forming a dense region around the observed measurement.

\textbf{After the second update ($k=2$):} When the next measurement $z_2 = 0.6$ arrives, the new evidence pulls the cloud back toward the left. After resampling, multiple particles cluster near $x \approx 1.5$, while a few remain farther out to reflect uncertainty. This dynamic movement of the particle cloud embodies the Bayesian trade-off between prediction and correction.

\section{Code Lab 1: Python Implementation}

To make the numerical example fully reproducible, let us implement the
one-dimensional particle filter in Python.
The following code follows exactly the same steps described in this chapter:
prediction, weighting, normalization, and resampling.

\begin{lstlisting}[language=Python, caption={Particle Filter for 1D Tracking (Code Lab 1)}]
import numpy as np
import matplotlib.pyplot as plt

# -----------------------------
# Simulation parameters
# -----------------------------
T = 15           # number of time steps
Q = 1.0          # process noise variance
R = 4.0          # measurement noise variance
N = 200          # number of particles

# True and observed values
x_true = np.zeros(T)
z_meas = np.zeros(T)

# Initialize true state
x_true[0] = 0.0

# -----------------------------
# Particle filter initialization
# -----------------------------
particles = np.random.normal(0.0, 2.0, N)  # prior p(x0) = N(0, 2^2)
weights = np.ones(N) / N

# Storage for state estimates
x_est = np.zeros(T)
x_est[0] = np.average(particles, weights=weights)

# -----------------------------
# Main loop
# -----------------------------
for k in range(1, T):
    # True system evolution
    eta_k = np.random.normal(0.0, np.sqrt(Q))   # process noise
    v_k = np.random.normal(0.0, np.sqrt(R))     # measurement noise
    x_true[k] = x_true[k-1] + eta_k
    z_meas[k] = x_true[k] + v_k

    # --- Prediction step ---
    particles += np.random.normal(0.0, np.sqrt(Q), N)

    # --- Weighting step ---
    likelihoods = np.exp(-0.5 * ((z_meas[k] - particles)**2) / R)
    weights *= likelihoods
    weights += 1.e-300      # avoid round-off to zero
    weights /= np.sum(weights)

    # --- Compute effective sample size (ESS) ---
    Neff = 1. / np.sum(weights**2)

    # --- Resampling step (systematic) ---
    if Neff < N / 2:
        # Systematic resampling
        positions = (np.arange(N) + np.random.rand()) / N
        indexes = np.zeros(N, 'i')
        cumulative_sum = np.cumsum(weights)
        i, j = 0, 0
        while i < N:
            if positions[i] < cumulative_sum[j]:
                indexes[i] = j
                i += 1
            else:
                j += 1
        particles = particles[indexes]
        weights.fill(1.0 / N)

    # --- State estimation ---
    x_est[k] = np.average(particles, weights=weights)

# -----------------------------
# Visualization
# -----------------------------
plt.figure(figsize=(8,4))
plt.plot(x_true, 'k-', label='True position')
plt.plot(z_meas, 'rx', label='Measurements')
plt.plot(x_est, 'b--', label='Particle filter estimate')
plt.xlabel('Time step k')
plt.ylabel('Position')
plt.legend()
plt.title('1D Particle Filter Tracking Example')
plt.tight_layout()
plt.show()
\end{lstlisting}

\subsection*{Explanation of key components}
\begin{itemize}
    \item \texttt{particles}: array of $N$ hypotheses about the state $x_k$.
    \item \texttt{weights}: importance of each particle based on measurement likelihood.
    \item \texttt{prediction step}: adds process noise $\eta_k$ to each particle.
    \item \texttt{weighting step}: compares predicted measurement to actual $z_k$.
    \item \texttt{resampling}: performed adaptively when the effective sample size falls below $N/2$.
    \item \texttt{x\_est[k]}: weighted mean representing the state estimate $\hat{x}_k$.
\end{itemize}

Running this code produces three curves:
the true position, the noisy measurements, and the particle filter’s estimated trajectory.
Even with relatively few particles, the estimate will closely track the true state
while maintaining robustness against measurement noise.

\subsection*{Discussion}

This small experiment demonstrates the full cycle of the particle filter in action:
\begin{enumerate}
    \item The particle cloud spreads during prediction,
          reflecting process uncertainty.
    \item It tightens around plausible regions during weighting and resampling.
    \item The resulting mean follows the true state closely, smoothing out noise.
\end{enumerate}

In later chapters, we will extend this code to higher-dimensional systems
and explore advanced resampling strategies, but this simple Python example
already captures the essence of the particle filter.

\chapter{Particle Filter in Higher Dimensions}

\section{State Vector Formulation}

Until now, we have focused on a one-dimensional example where the state consisted of a single quantity—position. In real-world problems, however, the state is almost always a vector that contains multiple quantities describing the system.

\subsection{General form}
We now define the state vector at time step $k$ as:
\[
x_k =
\begin{bmatrix}
[x_k]_1 \\ [x_k]_2 \\ \vdots \\ [x_k]_n
\end{bmatrix}
\in \mathbb{R}^n,
\]
where $n$ is the dimension of the state space and $[x_k]_j$ denotes the $j$-th component of $x_k$.

For example, in a 2D tracking problem:
\[
x_k =
\begin{bmatrix}
p_x \\ p_y \\ v_x \\ v_y
\end{bmatrix},
\]
where $p_x, p_y$ are position components and $v_x, v_y$ are velocities. This compact representation allows the particle filter to operate in any number of dimensions without changing its basic logic.

\subsection{Particle representation in vector form}

Each particle now becomes a vector sample $x_k^{(i)} \in \mathbb{R}^n$ rather than a scalar. We maintain a set of $N$ such particles, each with an associated weight:
\[
\{x_k^{(i)},\, w_k^{(i)}\}_{i=1}^{N}, \quad
w_k^{(i)} \ge 0, \quad
\sum_{i=1}^{N} w_k^{(i)} = 1.
\]
The empirical approximation to the full posterior remains the same:
\[
p(x_k \mid z_{1:k})
\approx
\sum_{i=1}^{N} w_k^{(i)}\, \delta(x_k - x_k^{(i)}),
\]
but now $x_k$ and $x_k^{(i)}$ are vectors. In implementation, this means that the motion and measurement updates are performed for each component of the state vector, and each particle carries a complete hypothesis about the system's multidimensional state.

\subsection{Dimensionality and computational scaling}

Although the mathematics generalizes easily, the computational cost increases rapidly with the number of dimensions. For example:
\begin{itemize}
    \item In 1D, $N = 100$ particles may suffice.
    \item In 4D or 6D, thousands of particles may be needed for the same coverage of state space.
\end{itemize}
This phenomenon is known as the \textbf{curse of dimensionality}. We will be uploading a follow up article that will discuss strategies such as Rao–Blackwellization and adaptive sampling to mitigate this issue. Despite this, the underlying algorithm remains identical: each particle carries a full state hypothesis, and the same predict–weight–resample cycle applies in vector form.

\section{The Motion and Measurement Models}
A higher-dimensional particle filter still operates according to two fundamental models: the \textbf{motion model} (also called the process model) and the \textbf{measurement model} (or observation model).

\subsection{The motion model}
The motion model describes how the state evolves from one time step to the next:
\[
x_k = f(x_{k-1}, u_k) + \eta_k,
\]
where:
\begin{itemize}
    \item $x_k \in \mathbb{R}^n$ – the new state vector at time $k$,
    \item $u_k \in \mathbb{R}^m$ – the control input vector (optional),
    \item $f(\cdot)$ – the possibly nonlinear function describing system dynamics,
    \item $\eta_k \sim \mathcal{N}(0, Q)$ – the process noise vector with covariance matrix $Q$.
\end{itemize}

Each particle is propagated independently through this function:
\[
x_k^{(i)} = f(x_{k-1}^{(i)}, u_k) + \eta_k^{(i)}, \qquad
\eta_k^{(i)} \sim \mathcal{N}(0, Q).
\]
This step broadens the particle cloud according to the process noise, representing our uncertainty about the true motion.

\subsection{The measurement model}
The measurement model links the hidden state to the observable quantities. In vector form:
\[
z_k = h(x_k) + v_k,
\]
where:
\begin{itemize}
    \item $z_k \in \mathbb{R}^o$ – the measurement vector,
    \item $h(\cdot)$ – the (possibly nonlinear) function that maps the state to expected sensor readings,
    \item $v_k \sim \mathcal{N}(0, R)$ – the measurement noise vector with covariance matrix $R$.
\end{itemize}

Each particle predicts what the sensor would observe if the system were in that particle’s state:
\[
\hat{z}_k^{(i)} = h(x_k^{(i)}).
\]
The likelihood of the actual measurement $z_k$ given each particle's prediction is then computed as:
\[
p(z_k \mid x_k^{(i)}) =
\frac{1}{(2\pi)^{o/2} |R|^{1/2}}
\exp\!\left[
    -\frac{1}{2}
    (z_k - \hat{z}_k^{(i)})^\top R^{-1} (z_k - \hat{z}_k^{(i)})
\right].
\]
These likelihoods form the unnormalized weights for the update step.

\subsection{Practical interpretation}
The two models together describe the complete probabilistic behavior of the system:
\begin{itemize}
    \item The motion model defines how the state moves diffuses over time.
    \item The measurement model defines how observations constrain and correct those predictions.
\end{itemize}

The particle filter alternates between these two models at every time step. Regardless of dimensionality, the algorithm’s flow remains identical:
\[
\text{Predict} \rightarrow \text{Weight (via likelihood)} \rightarrow \text{Resample}.
\]

\subsection{Modeling considerations}

When extending to higher dimensions, careful modeling becomes essential:
\begin{itemize}
    \item The covariance matrices $Q$ and $R$ should reflect realistic coupling between variables.
    \item Nonlinear functions $f(\cdot)$ and $h(\cdot)$ must capture how the system actually behaves — not just approximate it linearly.
    \item For control inputs $u_k$, the model must encode how actions (like acceleration or steering) influence the next state.
\end{itemize}

With these foundations, the particle filter can handle any dimension of state and observation, from a simple 1D position estimate to a full 6D pose tracking system in robotics.

\section{2D Visualization of the Particle Cloud}
Visualizing a two-dimensional particle filter provides an intuitive picture of how the filter behaves in higher dimensions. Each particle now represents a point in the $(x, y)$ plane, rather than a scalar along a line. The distribution of these points across time shows how uncertainty moves and shrinks as the filter updates its belief.

\textbf{Initial belief:} At $k=0$, particles are sampled from the prior distribution $p(x_0)$. They may appear as a diffuse circular cloud centered around an initial guess, for example $(0, 0)$, reflecting uncertainty in both position coordinates. If velocity components are part of the state vector, their influence on motion is not yet visible, since the first step has not occurred.

\textbf{Prediction step:} When the motion model is applied, each particle moves according to its predicted dynamics:
\[
x_k^{(i)} = f(x_{k-1}^{(i)}, u_k) + \eta_k^{(i)}, \qquad \eta_k^{(i)} \sim \mathcal{N}(0, Q).
\]
In a 2D position–velocity model, this typically shifts the cloud in the direction of motion and slightly spreads it due to process noise. The resulting plot might show the entire swarm moving in a smooth trajectory, with light diffusion indicating growing uncertainty.

\textbf{Weighting step:} When a new measurement $z_k$ arrives (for example, a noisy $(x, y)$ position reading), particles are reweighted according to their likelihoods $p(z_k \mid x_k^{(i)})$. Graphically, this means that particles near the observed measurement brighten or enlarge (depending on the plotting style), while those far away fade. The particle cloud’s density now increases around regions consistent with the observation.

\textbf{Resampling step:} Resampling then eliminates low-weight particles and replicates high-weight ones. Visually, this appears as a tighter cluster of points around the high-likelihood region. Over successive time steps, the particle swarm follows the true trajectory of the system, adapting its shape to nonlinear motions or sudden measurement changes.

\section{Code Lab 1: 2D Particle Filter in Python}
We now extend the 1D implementation to a simple 2D tracking scenario. The object moves on a plane with position $(p_x, p_y)$ and velocity $(v_x, v_y)$. The state vector is
\[
x_k =
\begin{bmatrix}
p_x \\ p_y \\ v_x \\ v_y
\end{bmatrix},
\qquad
z_k =
\begin{bmatrix}
p_x \\ p_y
\end{bmatrix} + v_k,
\quad v_k \sim \mathcal{N}(0, R),
\]
and the motion model follows
\[
x_k = F x_{k-1} + \eta_k, \qquad
F =
\begin{bmatrix}
1 & 0 & \Delta t & 0 \\
0 & 1 & 0 & \Delta t \\
0 & 0 & 1 & 0 \\
0 & 0 & 0 & 1
\end{bmatrix}, \quad
\eta_k \sim \mathcal{N}(0, Q).
\]

\begin{lstlisting}[language=Python, caption={2D Particle Filter (Code Lab 1)}]
import numpy as np
import matplotlib.pyplot as plt

# ----------------------------------
# Simulation parameters
# ----------------------------------
T = 30             # time steps
dt = 1.0           # time interval
Q_pos = 0.2        # process noise (position)
Q_vel = 0.05       # process noise (velocity)
R_meas = 2.0       # measurement noise variance
N = 500            # number of particles

# ----------------------------------
# State initialization
# ----------------------------------
x_true = np.zeros((T, 4))   # [px, py, vx, vy]
z_meas = np.zeros((T, 2))
x_true[0] = [0, 0, 1, 0.5]  # initial position & velocity

# ----------------------------------
# Particle initialization
# ----------------------------------
particles = np.random.normal(0, 2.0, (N, 4))
weights = np.ones(N) / N

# ----------------------------------
# Helper: state transition function
# ----------------------------------
def f(x, dt):
    F = np.array([[1,0,dt,0],
                  [0,1,0,dt],
                  [0,0,1,0],
                  [0,0,0,1]])
    return (F @ x.T).T

# ----------------------------------
# Main filter loop
# ----------------------------------
for k in range(1, T):
    # True motion
    eta_k = np.random.multivariate_normal(
        np.zeros(4),
        np.diag([Q_pos, Q_pos, Q_vel, Q_vel])
    )
    x_true[k] = f(x_true[k-1], dt) + eta_k

    # Measurement
    v_k = np.random.multivariate_normal(
        np.zeros(2),
        R_meas * np.eye(2)
    )
    z_meas[k] = x_true[k, :2] + v_k

    # --- Prediction ---
    noise = np.random.multivariate_normal(
        np.zeros(4),
        np.diag([Q_pos, Q_pos, Q_vel, Q_vel]),
        N
    )
    particles = f(particles, dt) + noise

    # --- Weighting ---
    diff = z_meas[k] - particles[:, :2]
    likelihood = np.exp(-0.5 * np.sum(diff**2, axis=1) / R_meas)
    weights *= likelihood
    weights += 1.e-300
    weights /= np.sum(weights)

    # --- Compute effective sample size ---
    Neff = 1. / np.sum(weights**2)

    # --- Resampling (systematic) ---
    if Neff < N / 2:
        positions = (np.arange(N) + np.random.rand()) / N
        cumulative_sum = np.cumsum(weights)
        indexes = np.searchsorted(cumulative_sum, positions)
        particles = particles[indexes]
        weights.fill(1.0 / N)

# ----------------------------------
# Visualization
# ----------------------------------
plt.figure(figsize=(6,6))
plt.scatter(particles[:,0], particles[:,1], s=5, color='blue', alpha=0.3, label='Particles')
plt.plot(x_true[:,0], x_true[:,1], 'k-', linewidth=2, label='True trajectory')
plt.scatter(z_meas[:,0], z_meas[:,1], c='r', marker='x', label='Measurements')
plt.xlabel('x position')
plt.ylabel('y position')
plt.title('2D Particle Filter Tracking Example')
plt.legend()
plt.axis('equal')
plt.tight_layout()
plt.show()
\end{lstlisting}

\subsection*{Explanation of the implementation}

\begin{itemize}
    \item The state vector $x_k = [p_x,\, p_y,\, v_x,\, v_y]^\top$ includes both position and velocity.
    \item The transition matrix $F$ advances each particle’s position according to its velocity.
    \item Process noise $\eta_k$ perturbs both position and velocity.
    \item Measurements observe only $(p_x, p_y)$ with Gaussian noise.
    \item The particle cloud evolves in 2D, clustering near the true path while reflecting uncertainty.
\end{itemize}

\subsection*{What you will observe}

Running this code will show (1) the true trajectory (black line) curving smoothly across the plane, (2) noisy measurements (red crosses) scattered around the path. (3) the final particle cloud (blue points) forming a tight ellipse near the true state.

The plot demonstrates how a 2D particle filter tracks motion in both coordinates simultaneously, maintaining a probabilistic representation that adapts naturally to nonlinear dynamics and noisy measurements.

\bibliographystyle{unsrt}
\bibliography{main}


\end{document}